\documentclass[preprint2]{aastex}

\slugcomment{To appear in ApJ}

\shorttitle{{\it ISO} Spectra of R Coronae Borealis Stars. I.}
\shortauthors{Lambert et al.}

\begin{document}

\title{ {\it Infrared Space Observatory} Spectra of R Coronae Borealis Stars. I.
Emission Features in the Interval 3 - 25 microns\altaffilmark{1}}

\author{David L.\ Lambert}
\affil{Department of Astronomy; University of
Texas; Austin, TX 78712-1083}
\email{dll@astro.as.utexas.edu}

\author{N. Kameswara Rao}
\affil{Indian Institute of Astrophysics;
Bangalore,  560034 India}
\email{nkrao@iiap.ernet.in}

\author{Gajendra Pandey and Inese I. Ivans}
\affil{Department of Astronomy; University of
Texas; Austin, TX 78712-1083}
\email{pandey@astro.as.utexas.edu, iivans@astro.as.utexas.edu}

\altaffiltext{1}{Based on observations with {\it ISO}, an ESA project 
with instruments funded by ESA Member States (especially the PI countries: 
France, Germany, the Netherlands, and the United Kingdom) and with the participation of ISAS and NASA.}

\begin{abstract}

{\it Infrared Space Observatory} 3 - 25 $\mu$m spectra of the
R Coronae Borealis stars V854 Cen, R CrB, and RY Sgr are presented
and discussed. Sharp emission features coincident in wavelengths
with the well known Unidentified Emission Features are present in
the spectrum of V854 Cen but not of R CrB or RY Sgr. Since
V854 Cen is not particularly H-poor and has a 1000 times
more H than the other stars, the emission features are probably
from a carrier containing hydrogen. There is a correspondence between
the features and emission from laboratory samples of hydrogenated
amorphous carbon. A search for C$_{60}$ in emission or absorption
proved negative. Amorphous carbon particles account for the broad emission
features seen between 6 - 14 $\mu$m in the spectrum of each star. 

{\it Subject headings:} stars:circumstellar matter --
stars: variables: other (R\,Coronae Borealis)

\end{abstract}

\clearpage

\section{Introduction}

The rare class of peculiar stars known as the
 R Coronae Borealis stars (here, R CrBs) possess two very distinctive
characteristics: a propensity to fade at unpredictable times by up to about
8 magnitudes, and an F-G supergiant-like atmosphere that is
very H-deficient, He-rich, with considerable amounts of carbon.  The fading of the
visible light is often rapid with deep minima achieved in a few days to a few
weeks but the return to maximum light is generally slower occurring over a period of
several months. O'Keefe's (1939) suggestion that a R CrB fades when a cloud of
soot forms to obscure the photosphere has stood the test of time but the  
detailed identification of the dust particles and their mode and site of
formation remain open questions. 

Infrared photometry and low-resolution spectroscopy of R CrBs at maximum light
and in decline have revealed several key features of the dusty circumstellar
shells. Discovery of IR-excesses provided confirmation of the hypothesis that
dust grains were a major constituent of the circumstellar shells (Stein et al.
1969; Lee \& Feast 1969). Subsequent studies showed the dust to be at an
equivalent blackbody temperature of 500 - 1000 K and present around almost all 
R CrB stars (Feast \& Glass 1973; Feast et al. 1977; Glass 1978), but not
distributed in a spherically homogeneous shell
 (Forrest, Gillett, \& Stein 1972).
Later observations including measurements of optical polarization suggested
the dust may be ejected in a preferred plane (Stanford et al. 1988;
Rao \& Raveendran 1993; Clayton et al. 1997). 
Feast et al. (1977) showed that for RY\,Sgr, a 38 day pulsating R CrB star, 
the intensity of the 3.5 $\mu$m emission from the warm dust  varied with
the same period as the visual light showing that the dust was heated 
by starlight. This variation was shown to continue with no
significant decline in mean intensity when the star experienced a
deep visual decline. The mean intensity did, however, drop at times when 
the star remained at maximum light. This extended series of
IR observations show that the circumstellar shell is composed of
discrete clouds; should a cloud form along the line of sight to the
star, a decline is witnessed but the heating of the collection
of clouds comprising the circumstellar shell is little affected
because only a modest fraction of the stellar surface is
blocked from directly heating the shell.  {\it IRAS} observations revealed,
in addition to the warm dust detected from the ground, an
extended (`fossil') shell of cold (T $\sim 30$ K)
 dust  (Gillett et al. 1986;
Rao \& Nandy 1986; Walker 1986).

Attribution of emission
and absorption features in spectra of circumstellar shells
around C-rich objects has been contested. Two leading proposals vie
for identification of emission features:
 polycyclic
aromatic hydrocarbons (here, PAHs), and hydrogenated
amorphous carbon (here, HAC). Since
R CrBs are H-deficient to different degrees, observation and analysis of their
infrared emission features affords an opportunity to
investigate the role of hydrogen in formation of the
very large molecules or dust grains.
Hydrogen deficiency and other circumstances peculiar to
R CrB circumstellar shells may promote formation of
other species.

With the advent of the {\it Infrared Space Observatory} ({\it ISO}, Kessler 
et al. 1996), there came
an  opportunity to obtain  IR spectra of R CrB stars in and beyond the narrow
windows open to ground-based observers. 
In this paper we discuss and interpret spectra of three R CrBs obtained with
 {\it ISO}
over the wavelength region 2.5 - 45 $\mu$m. 
At the H-rich end of the range is V854 Cen with a H abundance only
 a factor of 100
below normal.
We contrast  V854 Cen with  RY Sgr and R CrB, two stars
with considerably less H than V854 Cen.

\section{Observations}

\subsection{The Stars}

Two stars were satisfactorily observed under our {\it ISO} program: RY Sgr and V854 Cen.
Data for R CrB were retrieved from the {\it ISO} archive.
 Details of the observations including the parameters TDT and AOT
are summarized in Table 1. 
At the time of observation, RY Sgr and R CrB
were  close to maximum
light according to the AAVSO light curves. 
 When observed V854 Cen appears to have
been recovering from minimum light: the V magnitude according to Lawson et al.  (1999)
was about 7.7 or  0.5 magnitudes below maximum light. 

\begin{deluxetable}{llccccc}
\tabletypesize{\scriptsize}
\tablewidth{0pt}
\tablecolumns{7}
\tablecaption{The SWS Observations}
\tablehead{
\colhead{Star}      & \multicolumn{2}{c}{Date\tablenotemark{a}} & \colhead{V\tablenotemark{b}} &
\colhead{TDT} & \colhead{AOT} & \colhead{Speed} \\
\cline{2-3} \\
\colhead{} & \colhead{UT} & \colhead{JD$^\prime$} & \colhead{} & \colhead{} & \colhead{} &
\colhead{} }
\startdata
V854 Cen & 1996 Sep 9 & 50336 & 7.7 & 29701401 & SWS01 & 4 \\
R~CrB    & 1998 Jan 15 & 50828 & 6.1 & 79200268 & SWS01 & 2 \\
RY~Sgr   & 1997 Mar 25 & 50533 & 6.6 & 49500503 & SWS01 & 4 \\
\enddata
\tablenotetext{a}{JD = JD$^\prime$ + 240000}
\tablenotetext{b}{See text}
\end{deluxetable}

\subsection{Spectra - Reduction and Calibration}

The spectra were obtained with the {\it Short-Wavelength Spectrometer}
 ({\it SWS},
de Graauw et al. 1996)
on the {\it ISO} spacecraft.
Spectra were recorded on separate 1 $\times$ 12 detector arrays. The
complete spectrum was recorded as 12 different but overlapping spectral
bands using different combinations of grating, aperture, and
detector array. Our spectra of RY Sgr and V854 Cen were observed
at the slowest scanning speed, and, therefore, the highest {\it SWS}
resolving power of $R = \lambda/\Delta\lambda \sim 1000$. The
archived R CrB spectrum had been taken at a higher speed and lower resolving
power ($R \sim 400$). 

Standard pipeline processing of the raw {\it SWS} data   
 involves
three principal steps:
subtraction of the dark current, assignment of the wavelength
scale, and the determination of the absolute fluxes. Dark current
measurements were obtained just prior and after an observation.
These were averaged and subtracted from the signal from the source.
Wavelengths were calculated from the recorded grating positions
and grating constants in the {\it SWS} database (Valentijn et al. 1996).
Similarly, the raw signals were converted to absolute fluxes (in Janskys)
using calibrations in the {\it SWS} database. Each of  the 12 bands was
reduced separately, and then combined to form a composite spectrum.

The output of the pipeline processing was
 a spectrum
with discontinuities across the edges between adjacent bands. 
Additional reduction procedures were then applied using the interactive
software
{\it ISAP} (Roelfsema et al. 1993).
Spectra in the 12 bands were grouped into 
four larger bands covering the wavelength regions:
2.35 - 4.1 $\mu$m, 4.0 - 12.0 $\mu$m, 12.0 - 28.0 $\mu$m, and 29.0 -
45.2 $\mu$m. A spectrum from a single detector was inspected
for abrupt jumps in signal level, and excessive noise levels.
Detectors so affected were dropped from the sample. A few bad data
points were edited out.
Following recommendations by Rita  Loidl (private communication), we
reduced the discontinuities by applying additive corrections relative to
the signal from a reference detector within each AOT band selected
based on two criteria: (1) it should provide the smallest discontinuity at the
band edges, and the slope that is most consistent with other detectors, and
(2) choice of the detector should not change the mean flux level.  
 The first
band (2.4 - 2.6 $\mu$m) was taken as the flux reference. We found
that band 7 (7 - 12.0 $\mu$m) showed a significant offset between the
scanning directions; the `down' scan spectra were always of higher flux than
the `up' scan spectra. This is attributed to `memory effects'; the
`down' scans were rejected.
This procedure provided spectra of higher signal to noise
ratio than achieved in all of our earlier attempts to
eliminate the discontinuities. 
Final spectra  after rebinning to  a resolving power of R = 1000 for
RY Sgr and V854 Cen and R=400 for R CrB are
shown in Figure 1 .

\begin{figure}
\plotone{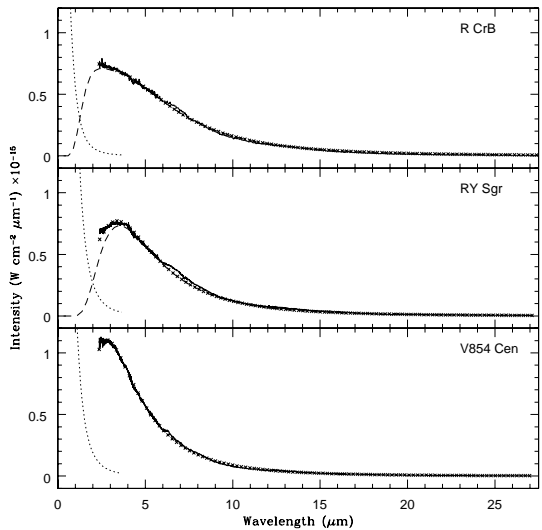}
\caption{{\it ISO SWS }spectra of R CrB, RY Sgr, and V854 Cen (solid lines).
The adopted photospheric continuum  is shown by the dotted
line in each panel. Assumed effective temperatures are 6900 K (R CrB), 7250 K
(RY Sgr), and 6750 K (V854 Cen). The dashed line shows for R CrB and
RY Sgr the blackbody continuum from the circumstellar dust for the
temperatures given in the text. The sum of the dust and photospheric
continuua is represented by the crosses. In the case of V854 Cen, the
very small contribution of the photospheric continuum is neglected
and the crosses give the dust continuum. Excess emission over the
fitted fluxes (crosses) is more clearly displayed in
Figures 2 and 3. \label{fig1}}
\end{figure}

\subsection{Spectral Energy Distributions}

Since the stars are of variable flux in the infrared, we should
 not expect the
{\it ISO} fluxes to match exactly
 published fluxes.
V854 Cen was
caught by {\it ISO} in an  
unusually bright phase at IR wavelengths: for example,  the {\it ISO} flux 
density at 10 $\mu$m is 25 Jy while 15 Jy was found by {\it IRAS}.
{\it IRAS} fluxes for RY Sgr are 60 - 70\% higher than the {\it ISO} fluxes. 
R CrB was about 40\% brighter as observed by {\it ISO} than by {\it IRAS}.
Walker et al. (1996) observed R CrB with {\it ISO} when the star was
at minimum light or 7 magnitudes below maximum light. The {\it ISOPHOT}
instrument was used for photometry at 6 bandpasses from 60 $\mu$m to 200 $\mu$m,
and a low resolution spectrum was acquired from 2.5 $\mu$m to 5 $\mu$m and
5.8 $\mu$m to 11.6 $\mu$m at a resolving power of about 100. The {\it ISO}
photometry seems to show that R CrB was a factor of about two brighter
at  minimum than when measured with {\it IRAS}. The 2.5 $\mu$m to 11.6 $\mu$m
spectrum differs in shape but not average flux from ours: the {\it ISOPHOT}
spectrum shows less flux than the {\it SWS} spectrum for wavelengths shorter
than about 4 $\mu$m and longer than about 8.5 $\mu$m with
differences largest at the wavelength limits of the {\it ISOPHOT} spectrum.
Between 4 $\mu$m and 8.5 $\mu$m, the {\it ISOPHOT} spectrum has
the higher flux by up to about a factor of two. The {\it ISOPHOT}
spectrum cannot be well fitted by a blackbody spectrum. 

%
%
%

Infrared excesses  are conveniently
expressed using a fit of a blackbody spectrum after consideration of
the photospheric spectrum. This is an adequate
artifice here because we are primarily interested in the emission
and absorption features.
Predicted fluxes from model atmospheres  show that 
the IR photospheric flux is essentially identical to that of a black body at
the stellar effective temperature (Asplund et al. 1997a,b).
We resolve the observed spectrum
into  photospheric and  circumstellar components; the former
does not exceed a 10\%  contribution at the short wavelength end of the
{\it SWS} bandpass.  Then, we
fit the circumstellar spectrum with that of  a   blackbody. Figure 1 shows
the photospheric and circumstellar components with the observed spectra. 
The blackbody's 
contribution is subtracted to show more clearly the emission
and absorption features that may be present (see Figure 2, 3, and 4).
 Over the {\it SWS} spectrum, interstellar
reddening, slight at B and V, is negligible. 

\begin{figure}
\plotone{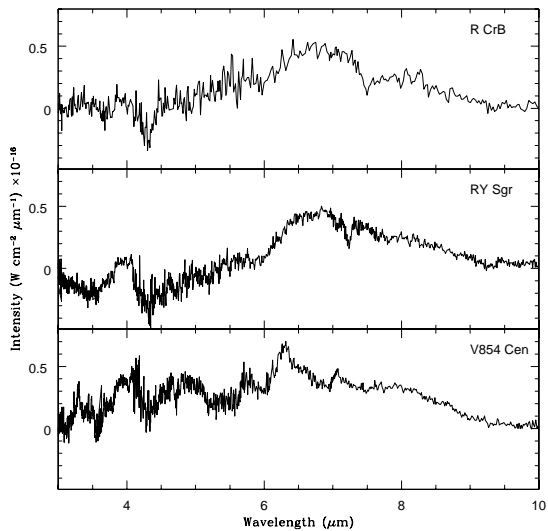}
\caption{Difference spectra of R CrB, RY Sgr, and V854 Cen from 3 to
10 $\mu$m. These are
the {\it ISO SWS }spectra after subtraction of the stellar
and circumstellar
continua, i.e., Figure 1's solid line minus the fitted distribution
represented by the crosses. Note the change of flux scale. \label{fig2}}
\end{figure}

\begin{figure}
\plotone{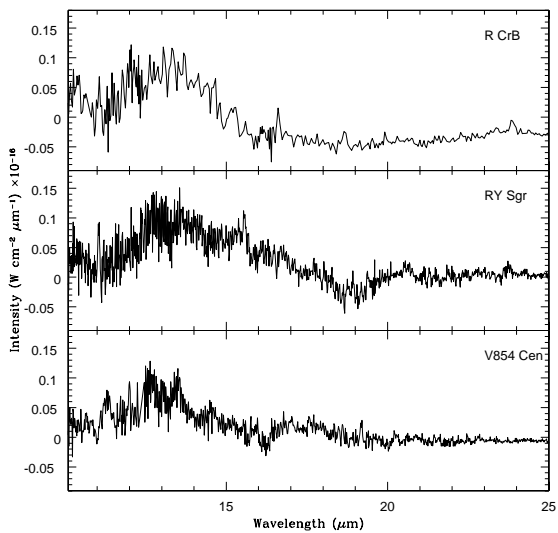}
\caption{Difference spectra of R CrB, RY Sgr, and V854 Cen from 10 to
25 $\mu$m. These are
the {\it ISO SWS }spectra after subtraction of the stellar
and circumstellar
continua, i.e., Figure 1's solid line minus the fitted distribution
represented by the crosses. Note the change of flux scale. \label{fig3}}
\end{figure}

For V854 Cen and RY Sgr, the infrared excess is well fitted by
a single blackbody: temperatures of 1040$\pm$20 and 820$\pm$10 K are
obtained for V854 Cen and RY Sgr, respectively. The former temperature is
slightly hotter than the 900 K estimated by Lawson \& Cottrell (1989)
from {\it IRAS} measurements, but is equal to the excitation temperature of
circumstellar C$_2$ molecules reported by Rao \& Lambert (2000). 
Our temperature for RY Sgr fits the {\it IRAS} measurements; Walker
(1985) derived a blackbody temperature of 800$\pm$50 K from
a fit to broad-band measurements out to 60 $\mu$m. A range of 
 600 K to 900 K was found by Feast et al. (1977) from
ground-based photometry. 
To fit the R CrB {\it ISO} spectrum requires two blackbody components:
 one at T=610 $\pm60$ K provides an
adequate fit at wavelengths longer than 5 $\mu$m, and the contribution from a
hotter black body, T = 1390$\pm$270 K,
 is needed to fit the shorter wavelengths.  Walker (1985) obtained a
temperature of 650$\pm$50 K from {\it IRAS} broad-band fluxes from 12 to 100
$\mu$m.
 We note that the {\it IRAS} observations suggested a similar
(T = 680 K) dust temperature (Rao \& Nandy 1986). 
In each case, the blackbody simulation of the {\it SWS} spectrum 
fits the observations
 to a few per cent with
small deviations largely arising from emission
features.

Our adoption of a blackbody to represent the infrared fluxes is
an artifice  enabling us to display more clearly the emission features.
Representation of an infrared flux distribution by a blackbody is
tantamount to assuming that either the dusty clouds are isothermal and optically
thick at all wavelengths  or the clouds are
optically thin with an absorption coefficient that is quasi-gray
in the infrared. An attempt to model the clouds as optically thin was made
using laboratory measurements of the non-gray absorption coefficient
for amorphous carbon samples
from Colangeli et al. (1995  - see also Bussoletti et al. 1987); the
measurements for their BE sample were adopted.
The inferred dust temperature
is slightly
 lower, and the overall
fit to the observed fluxes somewhat inferior to the results for
the optically thick case.
Presumably, the latter result
is, in part, due to the failure of the adopted laboratory
measured absorption coefficients to represent the actual circumstellar
dust; the laboratory results depend on how the amorphous carbon 
is prepared. 
More refined models than either our simple optically thick and thin
approximations would presumably
provide higher quality fits but our present purpose of investigating the
narrow emission features is served adequately by 
adopting the blackbody fit to the spectrum.

Amorphous carbon particles show increased absorption between about 6 $\mu$m
and 14 $\mu$m (see, for example, Colangeli et al.'s Figure 5, 1995)
 that appear to account for the broad emission
seen from all three R CrB stars. In the optically thick case, one expects
these features to appear in absorption on the grounds that the temperatures
of dust grains on clouds' outer edges will decrease towards the surface
facing away from the star. Optically thin clouds will show emission
bands associated with the increased absorption.
Colangeli et al.'s measurements, and the
temperatures found from the `optically thin' fit to the infrared
fluxes provide a rough fit to the broad emission features.

\section{Emission and Absorption Features}

\begin{figure}
\plotone{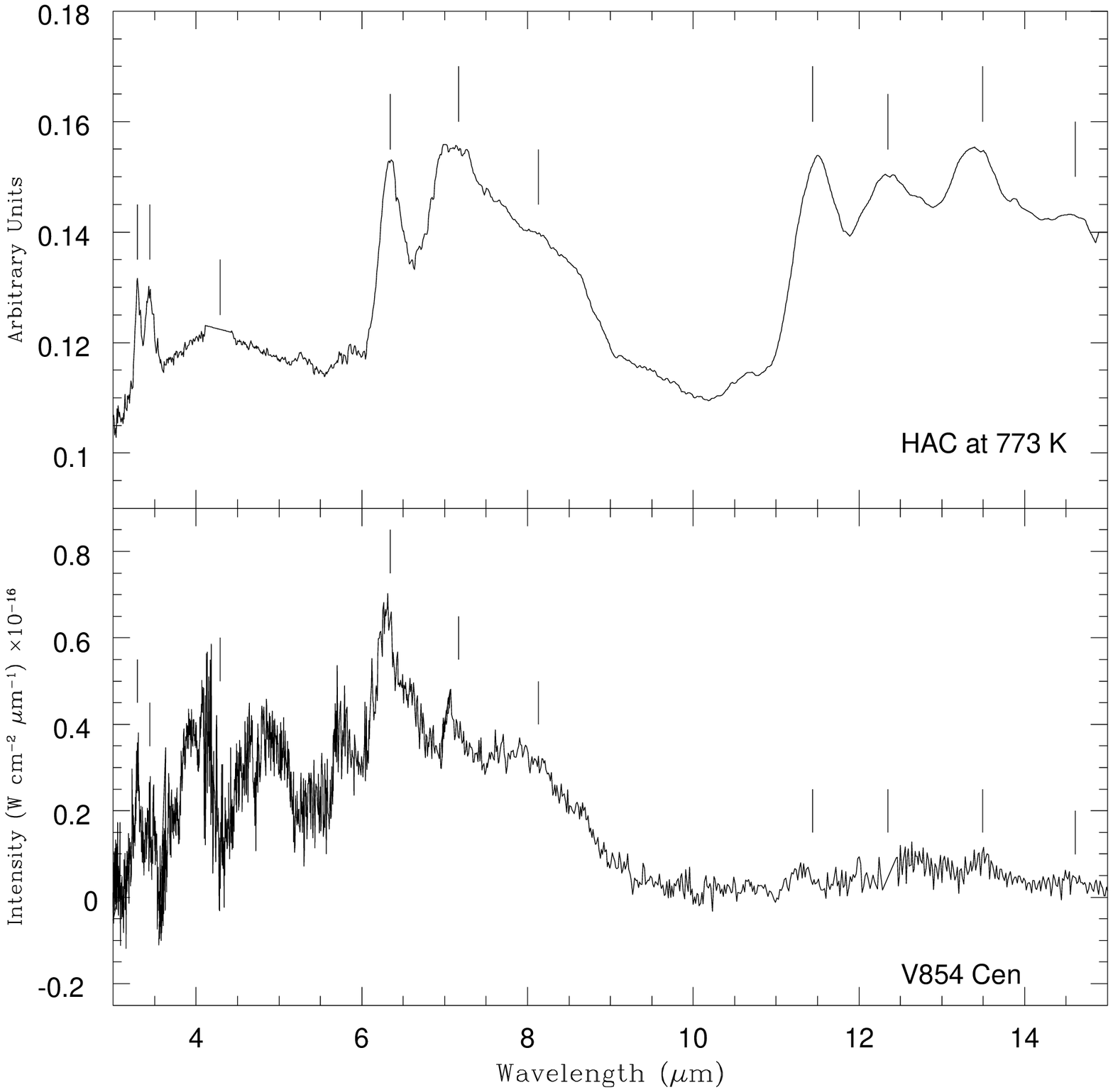}
\caption{Comparison of the difference spectrum of V854 Cen and the laboratory
emission spectrum of hydrogenated amorphous carbon at 773 K from
Scott et al. (1997). \label{fig4}}
\end{figure}

\subsection{Unidentified Infrared Emissions}

The principal goal of this investigation was to search for spectral features. 
Such features are seen in the residual or difference
 spectra (Figures 2,  and 3) obtained by subtracting the best-fitting blackbody
spectrum from the observed spectrum.
Emission features may be classified as `broad' or `narrow'. Narrow features
are seen only in V854 Cen. Broad features are seen in all three R CrB stars.

{\bf V854 Cen.}
The narrow emission features (Table 2) seen in V854 Cen's spectrum 
 correspond  to  some of the famous unidentified
infrared features (UIR) seen in emission from many post-AGB stars, planetary
nebulae, and H\,{\sc ii} regions. 
The 3.29 $\mu$m UIR is certainly present (Figure 2), 
the 3.4 $\mu$m UIR is possibly present, but the UIR at 3.51 $\mu$m is not
seen above the noise.
 In the 5-10 $\mu$m interval, the UIR
at 6.29 $\mu$m  but not those
at 5.6, 6.9,  7.3, 7.7, and 8.6 $\mu$m  stand out
above the noise and/or the smooth profile of the broad feature that
extends from 6-9 $\mu$m.
Three UIRs are seen in the 10-25 $\mu$m region (Figure 3): emission is
clearly seen at 11.3, and  13.5, and possibly 14.6 $\mu$m.  The latter two 
UIRs were discovered in {\it ISO} spectra of Red Rectangle (Waters et al. 1998).

 The remaining emission  features  are broad. 
A feature appears at 3.95 $\mu$m but it occurs at the border of two
spectral bands.
 The 6.3 $\mu$m UIR feature
appears at the short wavelength limit of
a broad feature extending to about 9 $\mu$m with local peaks
at 6.9 $\mu$m (possibly the intrusion of the 6.9 $\mu$m UIR)
 and 8.1 $\mu$m.
Clayton et al. (1995) observed V854 Cen from 8.5 to 8.8 $\mu$m at
a resolving power of 1000 and found a featureless continuum, an
observation consistent with our spectrum. 
An apparently broad  feature at 12.6 $\mu$m  
is bracketed by  UIRs at 11.3 $\mu$m and 13.5 $\mu$m. 
Inspection of Buss et al.'s (1993) collection of
6 - 13 $\mu$m  spectra of `transition objects'
(i.e., stars evolving from the AGB to the planetary nebula phase)
shows a gross similarity between V854 Cen and the C-rich protoplanetary
nebula {\it IRAS}  22272+5435.
 The spectrum of V854 Cen is definitely different from those
of the planetary nebulae NGC 7027 and CPD -56$^\circ$8032. 

\begin{deluxetable}{rccc}
\tabletypesize{\scriptsize}
\tablecolumns{5}
\tablewidth{0pt}
\tablecaption{V854 Cen - The Narrow Emission Features}
\tablehead{
\multicolumn{3}{c}{Parameters\tablenotemark{a}} & \colhead{} \\
\cline{1-3} \\
\colhead{$\lambda$} & \colhead{FWHM} & \colhead{Flux} & \colhead{} \\
\colhead{[$\mu$ m]} & \colhead{[$\mu$ m]} & \colhead{[10$^{-17}$ W cm$^{2}$]} & \colhead{UIR\tablenotemark{b}} }
\startdata
3.294(02) & 0.052(02) & 0.116(12) & Yes \\
6.298(10)   & 0.177(11)&0.378(70) & Yes \\
11.280(10)  & 0.443(11)&0.363(30) & Yes \\ 
13.490(17)    & 0.220(20) & 0.100(20)   & Yes \\
\enddata
\tablenotetext{a}{The estimated uncertainty is given in parantheses
following the entry, i.e., \\ 3.294(02) = 3.294 $\pm$ 0.002}
\tablenotetext{b}{See list of UIRs given by \\ http://www.ipac.caltech.edu/iso/lws/unidentified.html}
\end{deluxetable}

{\bf RY Sgr.}
Two broad features are present. One extends from about 6 $\mu$m to 9 $\mu$m,
and the second from about 11 $\mu$m to 18 $\mu$m. There is little evidence of
structure within these features. Except for the presence of the 6.3 $\mu$m
feature in V854 Cen, the appearance of these broad features is
similar for  V854 Cen and RY Sgr.   
Clayton et al. (1995)  obtained a  spectrum of
the interval 8.2 to 8.8 $\mu$m at a resolution of 1000 and reported that it
was featureless, a result consistent with our spectrum. The {\it IRAS LRS} spectrum
shown by Clayton et al. provides a marginal detection of the 11.5-15 $\mu$m
feature, and clear evidence for stronger emission extending from 
the 8 $\mu$m limit of the {\it LRS} spectrum to 9 $\mu$m, which we identify with the
6-9 $\mu$m broad emission feature.

{\bf R CrB.}
Again, two broad emissions are present.
We  confirm the 6 to 9 $\mu$m feature found by Buss et al. (1993),
and partially recorded on
 the {\it IRAS LRS} 8-22 $\mu$m spectrum (Clayton et al. 1995).
An emission peak at 6.5 $\mu$m 
in the feature reported by Buss et al. is not confirmed.
A second broad feature
extends from 11.5 to about 15 $\mu$m. 
Both features commence at the same wavelength in
all three stars but  span a shorter wavelength interval  in R CrB
than do the apparently
similar features in V854 Cen and RY Sgr.

Absorption near 4.6 $\mu$m seen in V854 Cen, and RY Sgr
is likely due to the fundamental vibration-rotation
absorption lines from circumstellar CO molecules.
Absorption at 7.2 $\mu$m in RY Sgr appears to be real; the
feature is present in each of the scans of this interval.

\subsection{A Search for C$_{60}$}

Goeres \& Sedlmayr (1992) may have been the first to have considered
the spherical cage molecule buckminsterfullerene, C$_{60}$,
as a possible constituent of dust clouds around R CrB stars. Their
theoretical model predicted low abundances of C$_{60}$ molecules. 
This  prediction  could be circumvented
if the carbon-rich gas contained H atoms. Formation of
large C-containing molecules is facilitated in the presence of H-containing
moleclues such as acetylene (C$_2$H$_2$) and by photoerosion of
hydrogenated amorphous carbon. This suggests that V854 Cen's dust clouds
might harbor C$_{60}$ molecules, even if RY Sgr and R CrB do not.
Understanding molecule and dust formation in the outer atmosphere
of a R CrB (or any star!) is far from complete. Therefore, a search
for these very stable carbon cages is of interest.

Laboratory infrared spectroscopy of free C$_{60}$ molecules  has
shown that  the strongest of the 46 possible vibrational bands  are
at 7.0, 8.4, 17.4, and 18.8 $\mu$m  
(Nemes et al. 1994, see also Kr\"{a}tschmer et al. 1990 and Frum et al.
1991). The given wavelengths are laboratory measurements extrapolated
to a temperature of 0 K.
Positions (and profiles) of these bands will
vary with temperature of the gas as more and more levels of the rotational
ladder are populated with increasing temperature.  Nemes et al.
estimate, for example, that the 8.4 $\mu$m will show its bandhead at 8.58 $\mu$m
at 1000 K
with a width of 0.1 $\mu$m.
Possibly, some molecules may be ionized - positively or negatively - in
the dust clouds.  Infrared spectroscopy of C$_{60}^+$ molecules in
a rare gas  matrix has provided measurements of  vibrational
bands of the ion C$_{60}^+$ at 7.1 and 7.5 $\mu$m (Fulara, Jakobi, \& Maier
1993); expected  bands at
longer wavelengths were not
investigated. The negative ion C$_{60}^-$ had strong bands at
 7.2 and 8.3 $\mu$m. 
These measurements include a matrix-dependent shift but the bands of the
gas phase
ions will be close to the measured positions. Temperature dependence of
 the band positions and widths will occur, as they do for C$_{60}$,
for the  free cations and anions.

Clayton et al. (1995) searched unsuccessfully for the C$_{60}$ 8.6 $\mu$m
band on ground-based R = 1000 spectra of the interval 8.49 $\mu$m to 8.82 $\mu$m
at a S/N ratio of about 100 for
R CrB but somewhat lower for V854 Cen and RY Sgr; the wavelength of
8.6 $\mu$m is that expected for C$_{60}$ absorption at 1000 K.
 We confirm 
that the feature is not present. (The S/N
ratio at 8.5 $\mu$m is 60 (V854 Cen), 70 (RY Sgr), and 80 (R CrB).) 
Our {\it SWS} spectra permit a  search for the other strong bands.
A feature appears near the 7.1 $\mu$m band in V854 Cen but this is at
the boundary of two {\it SWS} bands and, therefore, we doubt that it is a real
emission feature. This wavelength is also close to  that of
bands measured for matrix-isolated C$_{60}^+$ and C$_{60}^-$. 
Other bands of these ions
  at 7.5 $\mu$m for C$_{60}^+$ and 8.3 $\mu$m for C$_{60}^-$
are not present in our spectra.
At longer wavelengths (Figure 3), we do not detect the 
C$_{60}$ 17.5 and 19.0 $\mu$m bands. An apparent emission feature at 19.2 $\mu$m
is an artefact.

\subsection{The [Ne\,{\sc ii}] 12.8 $\mu$m Line}

Optical spectra of R CrBs in deep declines show a few broad emission
lines with base widths of 400 - 600 km s$^{-1}$.  Carriers of permitted
broad lines include 
 He\,{\sc i}, Na\,{\sc i} D, 
Ca\,{\sc ii}, and the C$_2$ molecule. Forbidden broad lines of
[N\,{\sc ii}], [O\,{\sc ii}] and
other atoms and ions have been noted.
 Detailed reports for V854 Cen (Rao \& Lambert 1993) and R CrB
(Rao et al. 1999) list and discuss the detected lines. Broad lines, if present,
would be resolved on the {\it SWS} R = 1000 spectra.
Broad lines,
which are detectable only after a star has faded by several magnitudes,
are not to be confused with sharp `chromospheric' lines,
primarily from neutral and singly-ionized metals, seen early in and
throughout a decline. Sharp lines would not be resolved on our spectra. 

Gas emitting the optical broad
 forbidden lines should also emit the [Ne\,{\sc ii}] 12.8 $\mu$m
line.
If the gas is of
low electron density, as indicated for V854 Cen by the intensity ratio of the
red [S\,{\sc ii}] lines (Rao \& Lambert 1993), the predicted flux of the
[Ne\,{\sc ii}] line is about 0.3\% that of the 6584\AA\ [N\,{\sc ii}]
line for equal abundances of the ions N$^+$ and Ne$^+$. Considerations of
elemental abundance and rough estimates of the ionization conditions
suggest that
the predicted flux of the
[Ne\,{\sc ii}] line is several orders of magnitude
smaller than the detection limit. Optical [O\,{\sc i}] and [C\,{\sc i}]
lines observed in spectra of R CrB at minimum light indicate electron densities
in excess of the critical densities for [N\,{\sc ii}] and [Ne\,{\sc ii}]
lines. Then, 
 the fluxes in the
[N\,{\sc ii}] and [Ne\,{\sc ii}] lines are comparable for equal ionic
densities, but, even in this more optimistic case, the flux of the
12.8 $\mu$m line
is at last a factor of 100 smaller than the detection limit. We conclude that
the absence of the 12.8 $\mu$m [Ne\,{\sc ii}] line is consistent with
optical detections of forbidden lines.

\section{Discussion}

\subsection{Hydrogenated Amorphous Carbon Dust}

Surely, the most intriguing result from these {\it ISO} spectra is the
presence of some UIRs in V854 Cen but not in RY Sgr and
R CrB. One may suspect this difference is related to the fact that
V854 Cen is less H-deficient than the other two stars.
 Then,
those UIRs seen in V854 Cen likely arise from transitions
involving H and C bonds in free molecules, large clusters, or
grains.

Analyses of the UV extinction by R CrBs suggested that the dust was amorphous
carbon (Hecht et al. 1984). Laboratory studies of absorption and emission
infrared spectra of aggregates of submicron particles or thin films of
partially hydrogenated amorphous carbon  have
been reported with a view to identifying the UIRs (Borghesi, Bussoletti,
\& Colangeli 1987; Scott \& Duley 1996; Scott, Duley, \& Jahani 1997).
In many cases, the precise profile of the absorption/emission feature
depends on the thermal history of the laboratory sample. Nonetheless,
there is a  correspondence between the narrow emission
features seen in the spectrum of V854 Cen and the laboratory spectra of HAC.
This is well illustrated in Figure 4 where the emission spectrum of a thin
HAC film at 773 K (Scott et al. 1997) is plotted with our spectrum of V854 Cen.
 The temperature of 773 K was
 the highest temperature investigated by Scott et al.
A majority of the emission features seen in the laboratory spectrum are
identifiable in V854 Cen's spectrum: absorption at 4.3 $\mu$m in V854 Cen
presumably masks the emission seen in the laboratory spectrum at a similar
wavelength, and the laboratory emissions at 7.1 $\mu$m and 12.3 $\mu$m are
  not seen as  distinct
peaks in the stellar spectrum. Apart from differences in the laboratory
and stellar profiles of the 7-9 $\mu$m and 11-15 $\mu$m bands, there is
an obvious difference in the relative strengths of these bands: the ratio
of the shorter to the longer wavelength emission is much greater
for the stellar than for the laboratory spectrum.

In the laboratory experiments, the intensity of the 3.4 $\mu$m feature,
characteristic of aliphatic hydrocarbons, decreased  as the temperature
was raised from 425 K to 775 K and the intensity of the 3.29 $\mu$m feature,
attributable to aromatic hydrocarbon and seen in V854 Cen, appeared and strengthened. 
This change was attributed by Scott et al. to the transformation of HAC from
a polymer to a protographitic solid.
In V854 Cen, the dominant emission is at about 3.3 $\mu$m not at 3.4 $\mu$m.
 The 6.2 $\mu$m (aromatic C-C ring vibration)
and 11.3 $\mu$m (aromatic CH-bending mode) are also more intense at the higher
temperatures. Emission corresponding to the CH bending modes is also apparent
at 12.3 and 13.2 $\mu$m in the 773 K laboratory spectra, and in V854 Cen.
 In short, these and
other comparisons of relative intensities suggest that a laboratory
spectrum at a temperature higher than 773 K, perhaps at the temperature of
1040 K of the dust shell, would prove to be an even closer
match to V854 Cen's narrow features.
UIRs seen in astronomical sources at 7.7 and 8.6 $\mu$m are apparently
absent from V854 Cen, and were also reported as absent by
Borghesi et al. (1987) in their studies of absorption spectra of
small HAC grains. The 7.7 $\mu$m feature was attributed to a CN-stretch
of a NH$_2$ group which owing to a lack of N in the laboratory samples
did not appear in laboratory spectra. Nitrogen is not especially
abundant in V854 Cen and, therefore, absence of this 7.7 $\mu$m UIR is
likely understandable.

RY Sgr and R CrB show  broad emission features  similar to those seen
from V854 Cen but the sharp features are exclusive to the latter
star. Profiles of the 7-9 $\mu$m and 11-15 $\mu$m bands differ from
star to star. Apart from the sharp features shown by V854 Cen, the
spectra of V854 Cen and R CrB are the most similar. The 7-9 $\mu$m band
appears resolvable into two bands with minimum flux at about 7.4-7.6 $\mu$m,
and the emission in the longer wavelength band does not extend beyond about
15 $\mu$m. By contrast, RY Sgr's 7-9 $\mu$m band does not show the
minimum, and emission in the longer wavelength band extends to about 18 $\mu$m.
These broad emission bands would appear to be a property of
dehydrogenated  amorphous
carbon (see Koike, Hasegawa, \& Manabe 1980; Bussoletti et al.
1987; Colangeli et al. 1995).

Hydrogenated amorphous carbon accounts well for V854 Cen's narrow emission
features.   Absence of strong features at 7.7 and 8.6 $\mu$m that are
characteristic of PAHs suggest the free molecules are not abundant
in its circumstellar shell. Amorphous carbon formed in a hydrogen
containing atmosphere is composed of randomly oriented linked and
connected PAH clusters, and, therefore, the spectrum of HAC will
resemble that of free PAHs with bands blurred, shifted, and blended
by solid state
effects. By an extension of this argument, it is not surprising that
an overall match to the HAC spectrum is also found from laboratory
spectrum of coals such as anthracite and semi-anthracite (Guillois et al.
1996; Papoular et al. 1996). We suppose it more probable that HAC rather
than coals form in V854 Cen's circumstellar shell.

\subsection{The Energy Budget}

It is of interest to investigate the energetics of the dust clouds 
which we assume are heated solely by absorption of starlight. 
Circumstellar dust is distributed in
clouds around the star.
After full recovery from a decline,
a star always returns to  the same brightness.  Therefore,
it is plausible to suppose that at these times there is
no cloud along the line of sight.
Although the detailed comparisons are lacking,
available comparisons of observed (corrected for interstellar extinction)
and predicted  stellar flux distribution about visible wavelengths
(see Asplund et al.'s Figure 8 1997a for R CrB) suggest that circumstellar
extinction by small particles is not present at maximum light; large particles
producing gray extinction cannot be ruled out by this comparison.
Gray extinction from a changing cloud of large particles
would cause fluctuations in the brightness of a star at maximum light.
If large dust particles  are made in the cloud responsible for a decline,
they cannot   persist in the line of sight once spectroscopic manifestations
of a decline (i.e., sharp emission lines) have disappeared in the recovery to
maximum light.

Given these assumptions, there is a simple inequality that  must be
satisfied by the integrated infrared emission from the dust ($f^d$) and the
integrated stellar flux ($f^*$), where $f$ denotes flux received at the Earth.
The inequality is $f^d < f^*$ and recognizes that (i) the dust clouds may
be optically thin, and (ii) the clouds may not subtend 4$\pi$ of solid
angle at the star. The dust emission, $f^d$, is estimated as described above
from the measured spectrum after subtraction of a small contribution from
the star. Stellar emission, $f^*$, is estimated from
published UBVRIJ magnitudes at maximum light with estimates added for
the ultraviolet flux from {\it IUE} spectra and for flux beyond J.

With the {\it ISO} fluxes, we find $f^d/f^* \simeq 0.9$
for V854 Cen, and 0.4 for R CrB and RY Sgr.
The larger ratio for V854 Cen may well result from the fact that we observed the
star in the recovery from a deep decline when the infrared fluxes were
enhanced due to a contribution from the recently formed dust cloud
that caused the decline. This suggestion is consistent with
Forrest's  (1974)  observations of R CrB  showing that the
infrared flux increased after mininum light with $f^d/f^*$ increasing from
0.2 to 0.7, by his estimates.  
 Noting too that the {\it IRAS} fluxes
for RY Sgr were 60 - 70\% higher than the {\it ISO} fluxes,
 we suggest $f^d \sim 0.5f^*$ for these
three stars at maximum light between declines.
Near-equality of $f^d$ and $f^*$ implies that the ensemble of clouds
provides a covering factor approaching 4$\pi$, and that a majority of
the clouds are optically thick or almost so. 
 Given this result, it seems surprising that the stars
are visible!  Or to phrase this surprise in another way - are there
R CrB stars that have eluded discovery because they are embedded in
a very optically thick  conglomeration of clouds?

If the dust clouds are optically thin, the mass in amorphous carbon can be
estimated readily. If the mean absorption coefficient at 7 $\mu$m, the peak of
the dust emission, is taken from Colangeli et al. (1995), and the stellar
distance derived from the stellar flux at about 1 $\mu$m and an assumed stellar
radius of 100 solar radii, the dust masses (in solar masses)
 are 2 $\times 10^{-8}$,
4 $\times 10^{-8}$, and 3 $\times 10^{-8}$ for V854 Cen, R CrB, and RY Sgr, 
respectively.  These are probably underestimates by a modest factor.
At minimum light, a R CrB is approximately 7 magnitudes fainter than
at maximum light. If the obscuring cloud covers the surface uniformly,
an optical depth of about 6 at visual wavelengths exists. Given the 
approximately inverse wavelength dependence of the dust absorption
coefficient, the optical depth of the cloud at 8 $\mu$m is about 0.4. This
estimate for a fresh cloud is likely an overestimate for the
typical cloud in the dust shell.    
  
To build a more complete picture of the clouds, imposition of radiative
equilibrium is required. Hartmann \& Apruzese (1976) were able to fit
Forrest's (1974) infrared photometry for infrared maximum and minimum (both
corresponding to visual maximum) 
with dust distributed out to 300 to 500 stellar radii. Optical
depth at infrared wavelengths for the assumed dust distributions is
very small. The dust mass derived of about 5 $\times 10^{-7} M_\odot$
is larger than our estimate but this is possibly due to differences
in the adopted absorption coefficients. A recalculation of model circumstellar
shells of HAC grains in radiative equilibrium  would be of interest.

\section{Concluding Remarks}

Dust in a cloud is reluctant to betray its identity. Yet, the
{\it ISO} 3 - 25 $\mu$m spectra of the ensemble of dust
clouds around the stars  V854 Cen, R CrB, and RY Sgr offer a few
new clues. Sharp emission features coincident with certain of the
UIRs are present only in V854 Cen's spectrum, and are 
a fair replica of emission seen in laboratory spectra of hydrogenated
amorphous carbon. These associations are consistent with the
fact that V854 Cen, although H-poor relative to normal stars, is
H-rich by a factor of 1000 with respect to R CrB and RY Sgr. Spectra of
all three stars show a double-peaked broad emission feature between
6 $\mu$m and 14 $\mu$m that corresponds to a feature in the extinction
curve of amorphous carbon. In summary, amorphous carbon is a major
contributor to the infrared emission from R CrB and, in the case of
V854 Cen, the amorphous carbon is hydrogenated.

 We are especially grateful to
Rita Loidl for advice on how to reduce the {\it SWS} spectra.
 We thank B. Gustafsson and M.Asplund for their interest in this project,
and W. Duley for providing
the laboratory spectrum illustrated in Figure 4.
The {\it ISO} Spectral Analysis Package ({\it ISAP}) is a joint 
development by the {\it LWS} and {\it SWS} Instrument Teams and
Data Centers with CESR, IAS, IPAC, MPE, RAL, and SRO as
contributing institutions.
This research was supported in part by NASA through grants NAG5-3348 and
CITJ-961543.

%
%
%
%
%
%
%
%
%
%
\end{document}